\documentclass{article}
\usepackage[utf8]{inputenc} % allow utf-8 input
\usepackage[T1]{fontenc} 
\usepackage[numbers]{natbib}

\usepackage{hyperref}
\usepackage[symbol]{footmisc}
\usepackage{afterpage}

\usepackage{booktabs}       % professional-quality tables
\usepackage{amsfonts}       % blackboard math symbols
\usepackage{nicefrac}       % compact symbols for 1/2, etc.
\usepackage{microtype}      % microtypography

\usepackage[final]{neurips_2018}

\usepackage{graphicx}
\usepackage{xcolor}
\usepackage{amsmath}
\usepackage{amssymb}

\author{%
  Thomas Schnake\thanks{t.schnake[at]gmail.com}, 
  David Bauer\thanks{bauer[at]gfai.de} \\
  Graph Based Engineering Systems\\
  Gesellschaft zur Förderung angewandter Informatik (GFaI)\\
  Berlin-Adlershof, 12489 \\
  \\
  2016
}

\title{Synthesis of High-Resolution Load Profiles with Minimal Data }
\date{2016}

\begin{document}
\newgeometry{top=2cm, left=4cm, right=4cm, bottom=3cm }

\begin{figure}[t]
    \includegraphics[width=4cm]{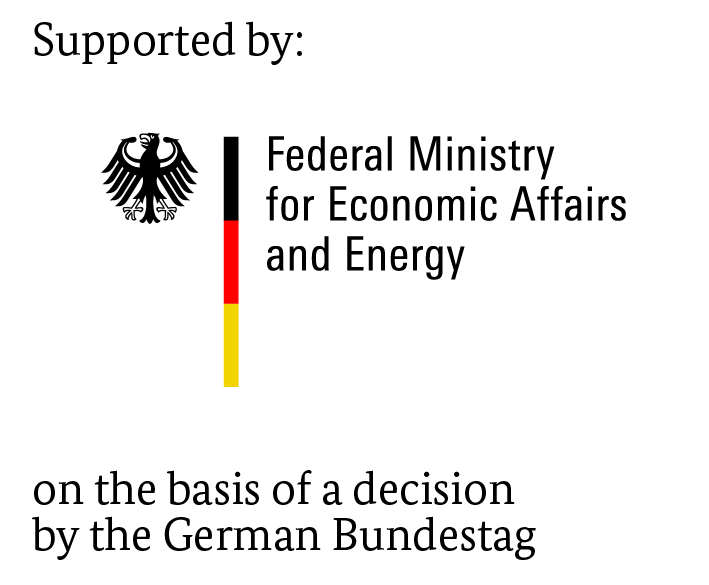}
     \hspace{5.6cm}
     \includegraphics[width=4cm]{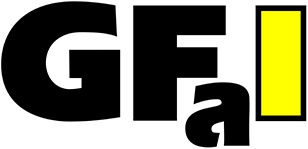}
\end{figure}

\maketitle

%\footnote[2]{ Gesellschaft zur Förderung angewandter Informatik (GFaI) \url{https://www.gfai.de/} }
%\begin{figure}
% 
%    Gesellschaft zur Förderung angewandter Informatik
%\end{figure}

\begin{figure}[ht]
    \centering
    \includegraphics[width=12cm]{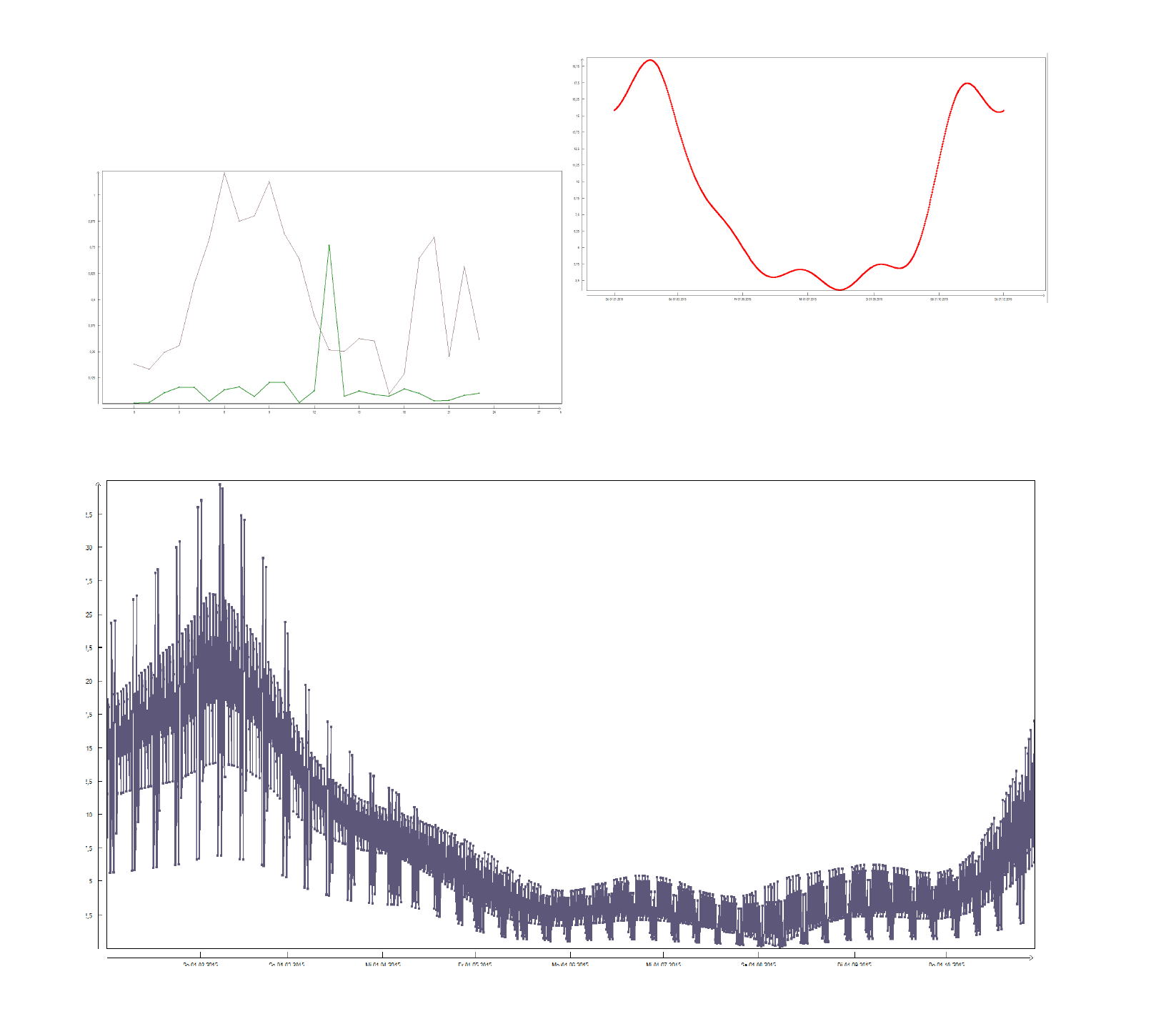}
\end{figure}

\thispagestyle{empty}
\newpage

\newgeometry{top=4cm, bottom=4cm, left=4cm, right=4cm }

\setcounter{page}{1}
\begin{abstract}
    For the estimation of of an new energy supply system it is an important to have high-resolution energy load profile. Such a profile is in general either not present or very costly to obtain. We will therefore present a method which synthesizes load profiles from minimal given data, but with maximal resolution. The general initial data setting includes month integrals and load profiles a few days. The resulting time series features all important properties to represent a real energy profile.
\end{abstract}

\section{Introduction}
In an energy supply system there often occurs the question of how to evaluate a version of an existing energy system. This means we know the costs and the energy need of the temporal state, but it is asked of the effect of the energy costs if we build for example another cogeneration plant. In this situation the simultaneity of energy and heat demand play a significant role, because the energy plants sometimes produce different forms of energy simultaneously. It is therefore important to estimate if and how much energy is demanded at the same time.\\
To be able to assess such system states it is important to model the load profiles as fine as possible to run a meaningful simulation. These load profiles do not exist in general in such a fine resolution, only in a coarse resolution such as month integrals \cite{epstein}, where in practice it is necessary to have a resolution per hour. We therefore want to synthesize a time series with a very fine resolution.\\

\section{The Data and the General Solution}
\begin{figure}[ht]
    \centering
    \includegraphics[width=12cm]{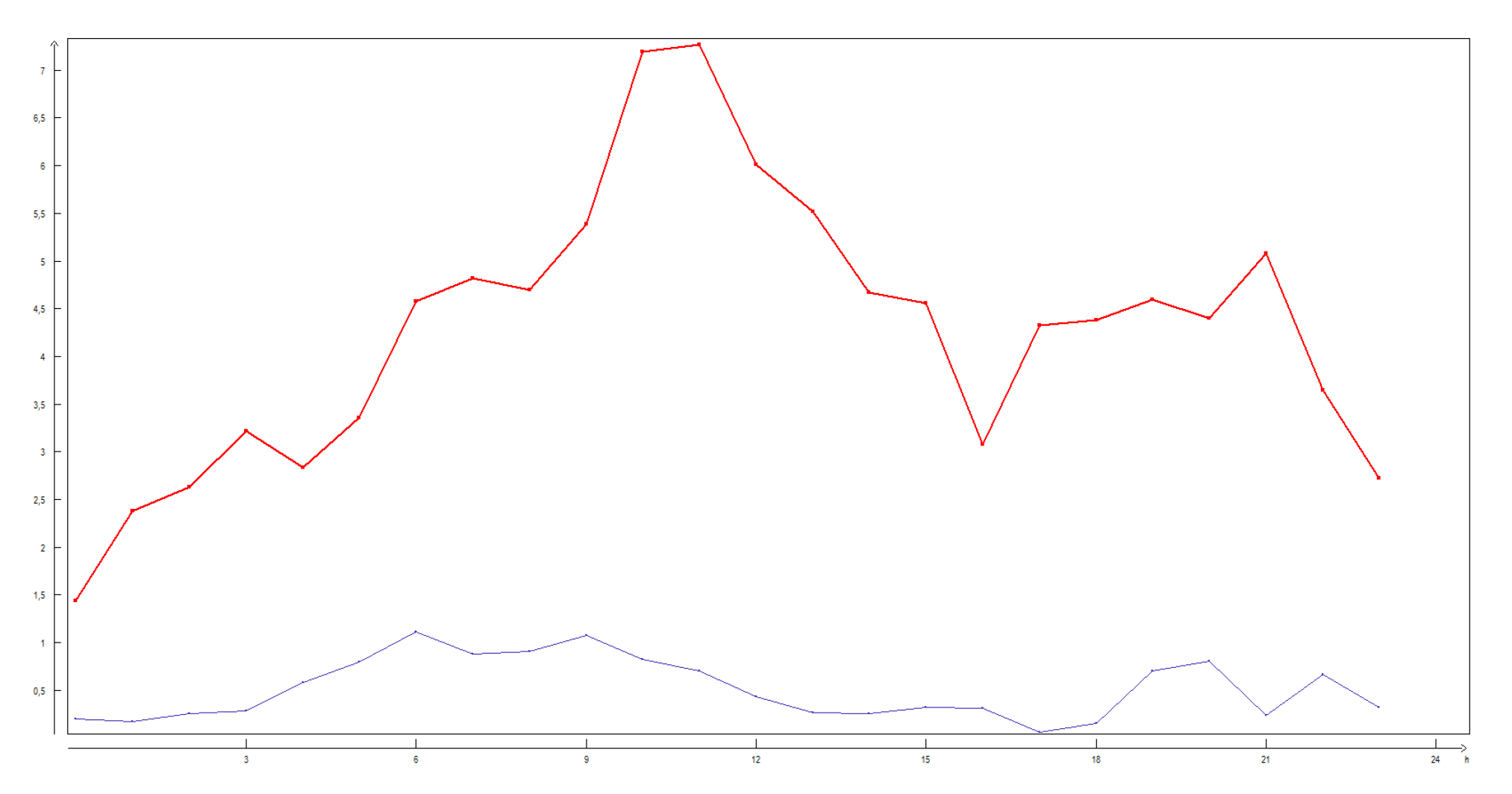}
    \caption{ Typical load profile of a one family home in summer (\textcolor{gray}{gray}) and winter (\textcolor{red}{red}) }
    \label{fig:dayloadsummerwinter}
\end{figure}

\begin{figure}
    \centering
    \includegraphics[width=12cm]{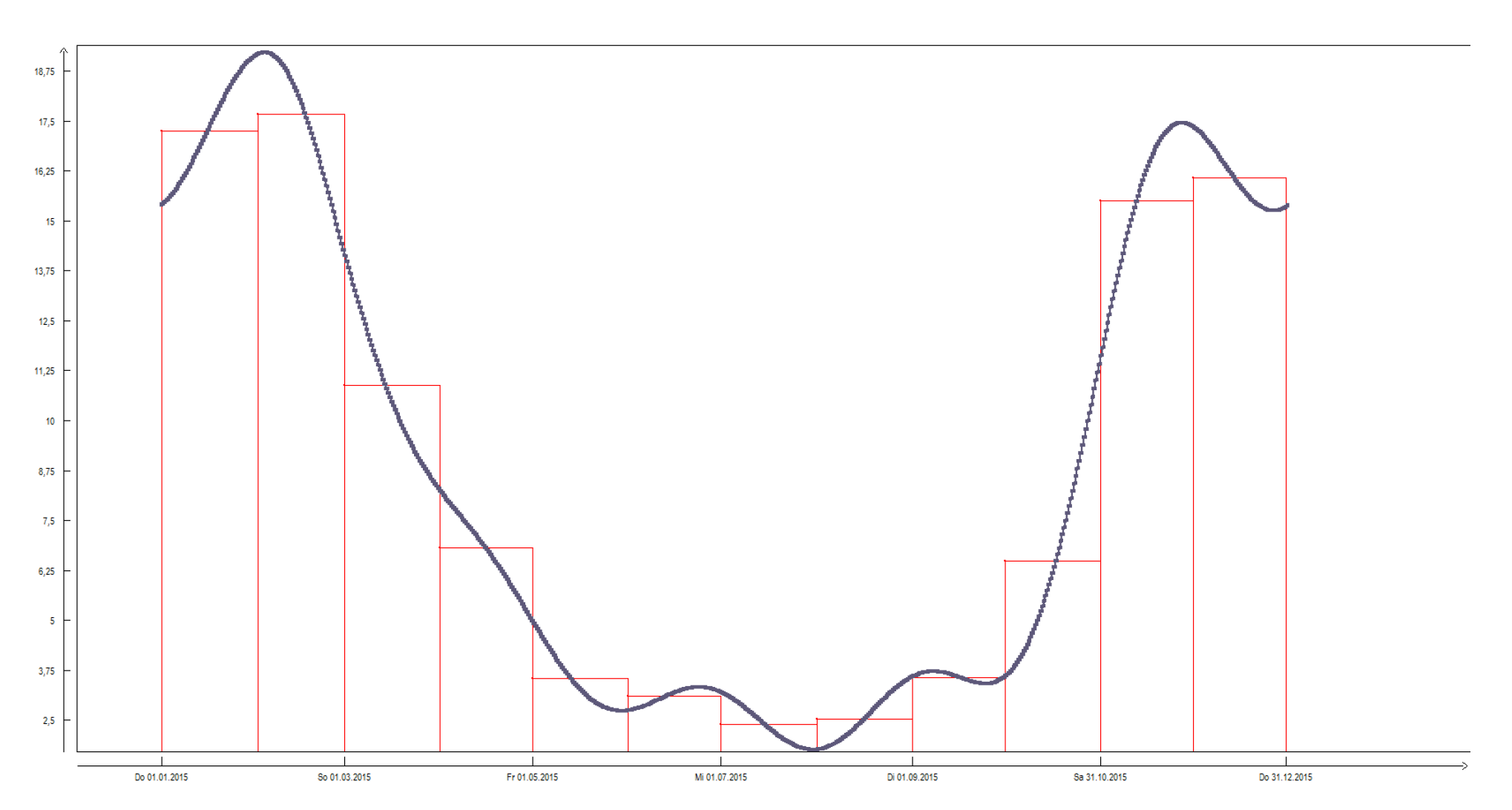}
    \caption{Example of 12 monthly integrals with coresponding interpolation function.}
    \label{fig:twelveMonthInterpol}
\end{figure}

In most situation one has 12 integral values for the estimated load profile given, one for each month. Now in order to extend the the profiles we take reference load profiles into account for each day in a week, which and each season of the year. First we want to construct with the 12 measurements a natural year structure. Figure \ref{fig:twelveMonthInterpol} shows an example of 12 month values transformed into a smooth interpolation curve. To obtain such a smooth curve from the 12 integral values we will use a interpolation method from E. Epstein \cite{epstein}.\\
Further we will discuss a semantic which provides with the few day profiles a natural structure for each day in the whole year. We will allow to have different day profiles within a week (e.g. energy profile in the week or weekend) and within the year seasons (e.g. heat requirement in winter and summer, figure \ref{fig:dayloadsummerwinter}). In particular we will be able to differentiate between seven different day profiles and four different season profiles per year. Since the day profiles will not change abruptly from one season to the other do a linear morphing of the day profiles within the year.\\ 
We want, with the help of given monthly integrals and day profiles, synthesize load profiles for the whole year which has relatively fine time resolution. For this it is important to maintain the integral for each month but still present the day profile in higher resolution, for example to keep extreme values.

\section{The Description of the Solution}
\subsection{Interpolation of the Monthly Integrals}
Let's consider we have given month integral values $(M_T)_{T=1}^{12}$. We want to discuss a reasonable interpolation of these integrals. A straigth forward approach would be to use linear or cubic splines \cite{epstein}, where we want to focus on maintaining the periodicity of the month integrals within the year.\\
We will take in particular results from the Fourier-Analysis and present the interpolation function as a finite 12-periodical Fourier series. Which means the interpolation function $y(t)$ is of the form
\begin{align}
    y(t) &= a_0 + \sum_{j=1}^N \left[ a_j \cos(\frac{2 \pi j t}{12}) + b_j \sin(\frac{2 \pi j t}{12} \right] & t \in [0,12]
\end{align}
where $N < \infty$ and $a_j, b_j \in \mathbb{R}$ for $j=0,...,N$. The time intervals $[T-\frac{1}{2} , T + \frac{1}{2} ]$ for $ T=1,..., 12$ are interpreted as the domain of the 12 month, which means that $y$ will take in the interval $[0.5, 1.5]$ values for the month January. \\
To determine the unknown coefficients $(a_j)_{j=0}^N$,  $(b_j)_{j=0}^N$ and $N$, we use the necessary condition that the integral over each month are given by $(M_T)_{T=1}^12$, which means
\begin{align}
    \int_{T-\frac{1}{2}}^{T+ \frac{1}{2}} y(t) \ dt = M_T \ \ \text{for } T=1,\dots, 12
\end{align}
It is verifiable that this problem can by solved with $N=6$ and the coefficients are given by
\begin{align*}
    a_0 &= \sum_{T=1}^{12} \frac{M_T}{12}\\
    a_j &= \left( \frac{\pi j}{12} \sin(\frac{\pi j}{12})^{-1}\right) \times
             \sum_{T=1}^{12} \left[ \frac{M_T}{6} \cos(\frac{2 \pi j T}{12}) \right] \ \ j=1,\dots, 5\\
     b_j &= \left( \frac{\pi j}{12} \sin(\frac{\pi j}{12})^{-1}\right) \times
             \sum_{T=1}^{12} \left[ \frac{M_T}{6} \sin(\frac{2 \pi j T}{12}) \right] \ \ j=1,\dots, 5\\
    a_6 &= \left( \frac{\pi}{2} \sin(\frac{\pi}{2})^{-1} \right) \times
            \sum_{T=1}^{12} \left[ \frac{M_T}{12} \cos(\pi T)\right]\\
    b_6 &= 0
\end{align*}
Figure \ref{fig:twelveMonthInterpol} shows an example of such an interpolation. Note that the the original integral values and the curve coincide exactly in the middle of the month.\\

\subsection{Seasonal Morphing of the Day Profile}
\begin{figure}
    \centering
    \includegraphics[width=12cm]{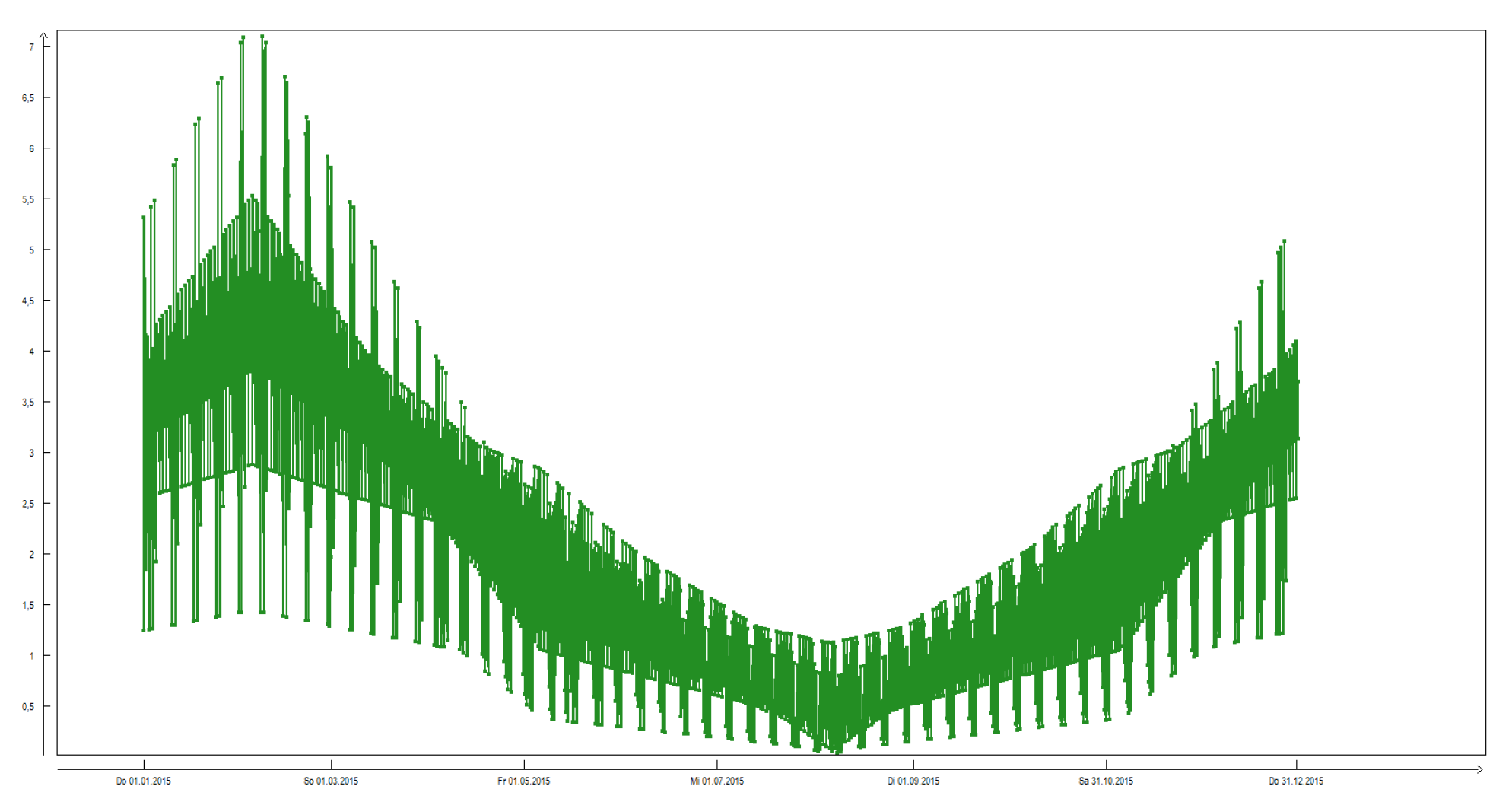}
    \caption{Picture of a stringing together of morphed day load profiles.}
    \label{fig:dayloadstringingmorphed}
\end{figure}

\begin{figure}
    \centering
    \includegraphics[width=12cm]{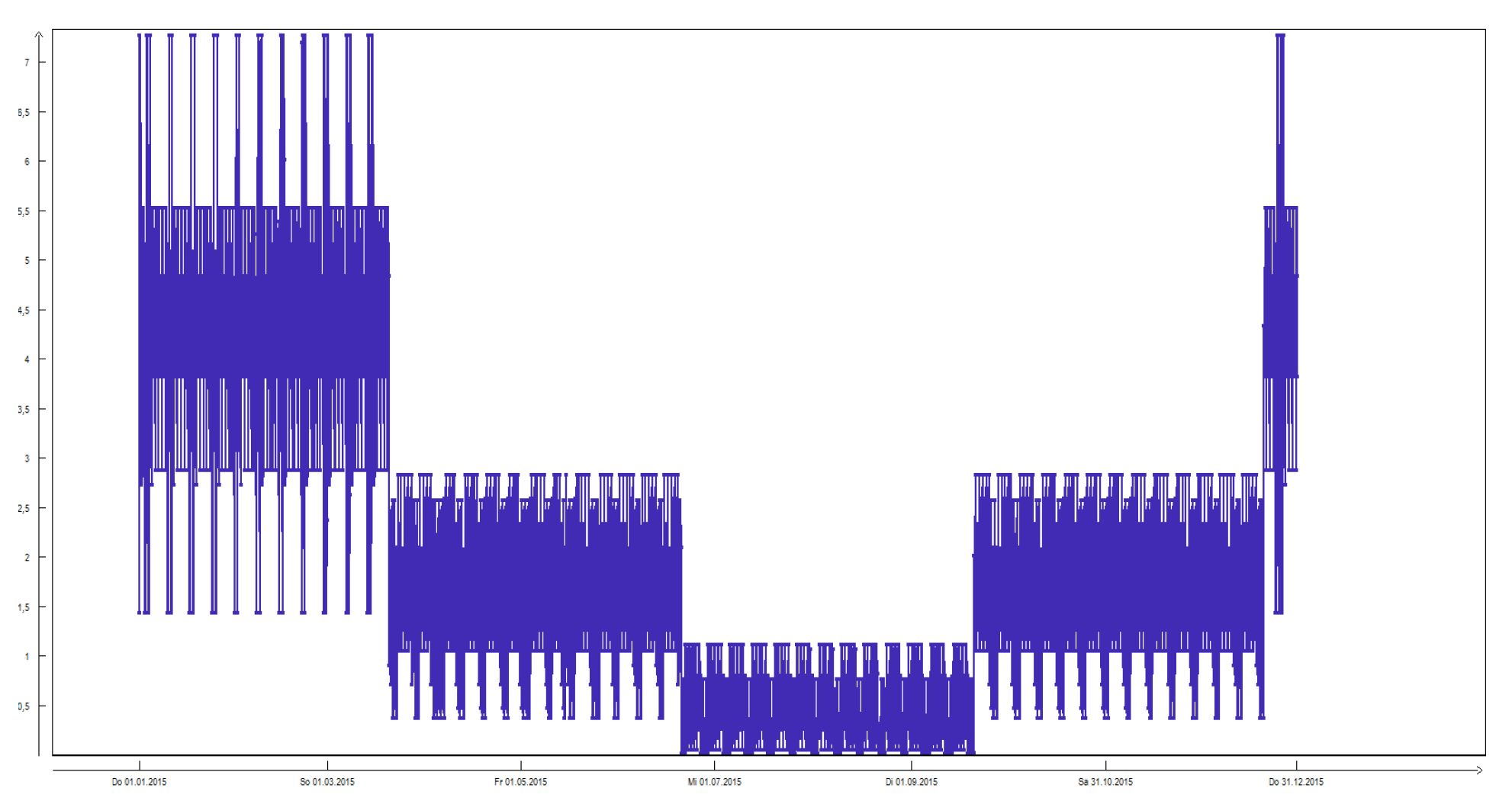}
    \caption{Picture of a stringing together of non morphed day load profiles.}
    \label{fig:dayloadstringing}
\end{figure}

We  differentiate between seven different type days per week times 4 different seasons in a year. These day profiles are given by the series $(\mathbb{T}_d^s)_{d,s}$ for $d=1,\dots, 7$ and $s=1,\dots,4$ where $d$ denotes the day and $s$ denotes the season. The reference load profile $\mathbb{T}_5^2$ will represent the load profile of a Friday in the second season of the year (e.g. Spring). For this case the load profiles differ in each season we want to ensure that there is no instant change at the transition from Spring to Summer with a smooth transition.\\
For an arbitrary season $s$ we denote time point $\mu_s$ to be the middle of the season $s$. We then say that the load profiles $(\mathbb{T}_d^s)_{d=1}^7$ is exactly the profile to obtain in the middle of $s$ i.e. at time point $\mu_s$. For the load profiles before or after the time point $\mu_s$ should be a mixture of the profiles $\mathbb{T}_d^s$ and $\mathbb{T}_d^{s-1}$ or $\mathbb{T}_d^s$ and $\mathbb{T}_d^{s+1}$ respectively.  The mixture is given by the a linear morphing of the type days in relation to the position between the two middle time points. This can be formalized as follows.\\
We demand for a reference load profile $\mathbb{T}^*$ at the date $t^*$, which is the weekday $d^*$ in the season $s^*$. Without loss of generality we consider $t^* \geq \mu_{s^*}$, because if $t^* < \mu_{s^*}$ we can just shift $s^* = s^* -1$ and obtain the same initial situation.\\
The linear morphing is then 
\begin{align}
    \mathbb{T} = \mathbb{T}_{d^*}^{s^*} + p(t*) \left[ \mathbb{T}^{s^* +1}_{d^*} - \mathbb{T}^{s^*}_{d^*} \right]
\end{align}
where
\begin{align}
    p(t) &= \frac{t-\mu_{s^*}}{ \mu_{s^* +1} - \mu_{s^*}}
\end{align}
for $t \in (\mu_{s^*}, \mu_{s^* +1})$. It is easy to see that
\begin{align}
    p(\mu_{s^*}) = 0, \ \ \ \ \  p(\mu_{s^* +1}) = 1
\end{align}
Figure \ref{fig:dayloadstringingmorphed} and \ref{fig:dayloadstringing} show the difference between morphed and non morphed load profiles, where the year integral is not yet considered.

\subsection{Composition of the Methods Under the Preservation of the Weekly Profile}
We want to connect the function $y(t)$ which we obtained by the interpolation of the month integrals, with the reference load profiles of each day to synthesize the yearly load profile $R(t)$. For this we also need a time series $w(t)$ which is given by stringing together the morphed type days, as in figure \ref{fig:dayloadstringingmorphed}. Let now $(t_i)_{i \in I}$ be the time grid of the year, i.e. the time points where $w$ is defined. We now want to define $R$ at $t_i$ with the property that $R$ keeps the same integral as $y$ on arbitrary intervals. More precisely we search for a coefficient $\alpha_i$ for $w(t_i)$ to obtain $R(t_i)$, which means
\begin{align}
    R(t_i) := \alpha_i w(t_i) \ \ \text{for } i \in I
\end{align}
We further want that $\alpha_i$ is sufficiently smooth, which means that $\alpha_i$ and $\alpha_{i+i}$ shouldn't vary too much. For this we assume that $w$ has a periodicity within a week so that for the time span of a week $W$ we expect that $\hat{w}(t) = \int_{t -W}^{t+W} w(s) \ ds$ is relatively smooth. Since the Fourier function $y$ is infinitely differentiable we expect that
\begin{align}
    \alpha_i := \frac{ \int_{t_i - W}^{t_i + W} y(s) \ ds }{\int_{t_i - W}^{t_i + W} w(s) \ ds }
\end{align}
meets the smoothness conditions. It is further possible to show that for arbitrary time intervals $\omega$ 
\begin{align}
    \int_{\omega} y(s) \ ds \approx  \int_{\omega} R(s) \ ds 
\end{align}

\section{An Example}
\begin{figure}
    \centering
    \includegraphics[width=12cm]{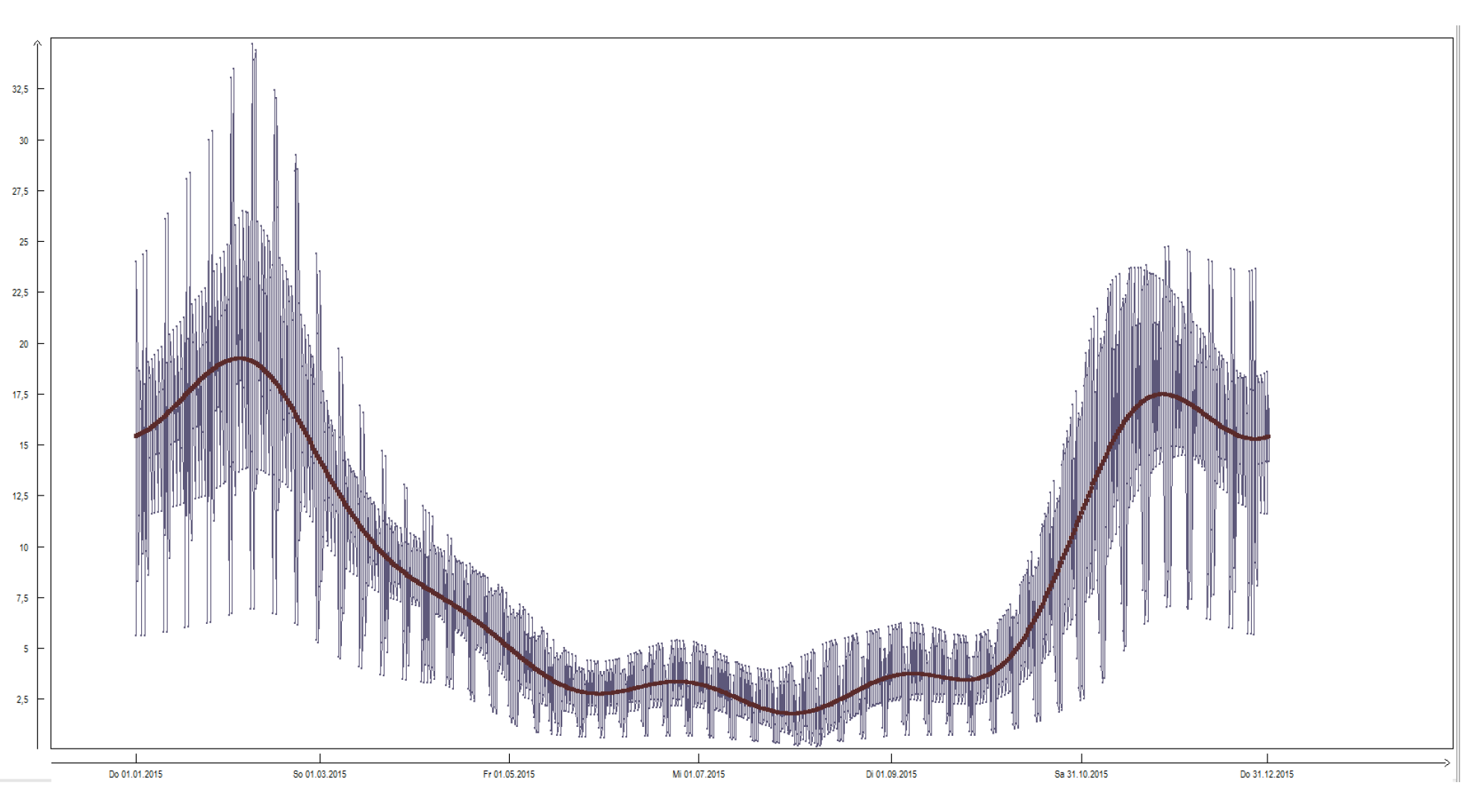}
    \caption{Picture of load profile for the whole year plotted against the interpolation function $y$.}
    \label{fig:yearprofilewithinter}
\end{figure}

\begin{figure}
    \centering
    \includegraphics[width=12cm]{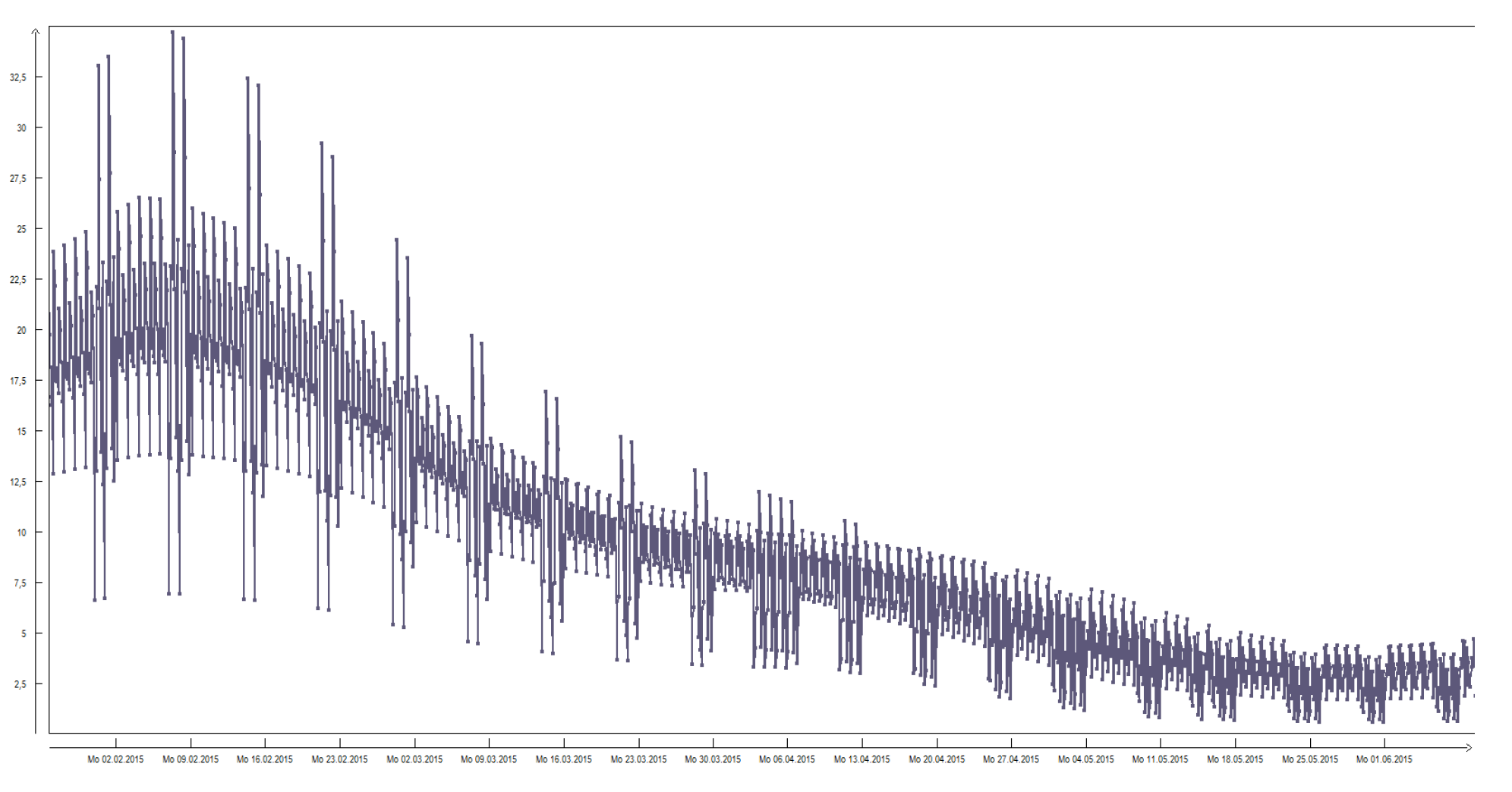}
    \caption{Picture of a closer look at the load profile to see the shape of the type days.}
    \label{fig:yearprofilecloserlook}
\end{figure}

We consider to have 12 month integrals of an appartement house given. Figure \ref{fig:twelveMonthInterpol} shows these integral values given with the corresponding interpolation function. Further we have type days for weekday and weekend in four different seasons, where in particular we chose the type days for Spring and Autumn to be the same. The type days where generated and published by VDI (engl. Association of German Engineers) 4655 \cite{VDI4655}. An example of such a load profile is visible in figure \ref{fig:dayloadsummerwinter}.\\
The beginning of each season is given by the meteorological seasons, i.e. the first of December, March, June and September for Winter, Spring, Summer and Autumn respectively. Figure \ref{fig:dayloadstringing} shows the day load profile stringing together with respect to each season.\\
In figure \ref{fig:yearprofilewithinter} and \ref{fig:yearprofilecloserlook} we see the the whole synthesized time series. In figure \ref{fig:yearprofilewithinter} we can also see the interpolation function of the month integrals from which it is good visible the the shape of it is inherited by onto $R$. Figure \ref{fig:yearprofilecloserlook} is a closer look on $R$, where we also see the fine resolution of the time series per day. 

\section{Conclusion}
It is possible to obtain with very few data a sufficient load profile to rate an energy system. The load profiles are not equal to recorded signal, but still exhibits the extreme points during the day and the integral for arbitrary time intervals. The obtained time series is very handy to obtain a reasonable estimations for new energy systems.

\section{Acknowledgement}
This research was done within the FuE-Project at Gesellschaft zur Förderung angewandter Informatik (engl. Society of the advancement of applied computer science) in Berlin-Adlershof. A special thanks goes to Stefan Kirschbaum for his helpful coorperation as well as to David Bauer and Gregor Wrobel.

\bibliography{bibliography.bib}
\end{document}